# The Northern ω-Scorpiid meteoroid stream: orbits and emission spectra

Francisco A. Espartero[1] and José M. Madiedo[2,1]

[1] Facultad de Ciencias Experimentales, Universidad de Huelva. 21071 Huelva (Spain).

[2] Departamento de Física Atómica, Molecular y Nuclear. Facultad de Física. Universidad de Sevilla. 41012 Sevilla, Spain.

**Corresponding author:** José M. Madiedo

Tel.: +34 959219991

Fax: +34 959219983

Email: madiedo@cica.es

**ABSTRACT**

We analyze the activity of the Northern ω-Scorpiid meteor shower between 2010 and 2012. For this purpose we have employed an array of low-lux CCD video cameras and spectrographs deployed at different astronomical observatories in Spain. As a result of our survey, the atmospheric trajectory and radiant position of 11 of these meteor events were obtained. The tensile strength and orbital parameters of the progenitor meteoroids have been also calculated. The calculated orbital data and the behaviour of these meteoroids in the atmosphere are compatible with an asteroidal origin of this stream. In addition, we discuss a unique emission spectrum recorded for one of these meteors. This is, to our knowledge, the first North ω-Scorpiid spectrum discussed in the scientific literature, and it has provided information about the chemical nature of the meteoroid and the progenitor body.

**1 INTRODUCTION**

Between the end of May and mid June, meteor and fireball activity from the Scorpiid-Sagitariid complex can be noticed. One of the streams belonging to this meteoroid complex is the North ω-Scorpiids, which was previously designated by some authors as the χ-Scorpiids (Jenniskens 2006). The North ω-Scorpiid shower is included in the IAU working list of meteor showers with the code 66 NSC (http://www.astro.amu.edu.pl/~jopek/MDC2007/). Its activity period goes from May 23 to June 15 with a maximum around June 1st. At its peak activity, its zenithal hourly rate (ZHR) is of around 5 (Kronk 1988). Previous observations of this shower were reported by Sekanina (Sekanina 1973, 1976) and Lindblad (Lindblad 1971a,b). The Apollo-type orbit found for NSC meteoroids let Drummond to propose 1862 Apollo as the parent asteroid of this meteor shower (Drummond 1981). However, this association was discarded, and nowadays Asteroid 1996 JG is included among the potential parent bodies of this stream (Jenniskens 2006).

In this paper we analyze a series of NSC meteor and fireball events recorded over Spain between 2010 and 2012. The orbit of the corresponding meteoroids was calculated and the tensile strength of these particles has been estimated. In addition the emission spectrum of a bright NSC meteor is presented and analyzed. This spectrum has provided some information about the likely nature of the progenitor body of the North ω-Scorpiids.

## **2 INSTRUMENTATION AND METHODS**

An array of low-light CCD video cameras (models 902H and 902H Ultimate, manufactured by Watec Co.) was employed to obtain meteor atmospheric trajectories and meteoroid orbits by triangulation. These devices operated at sites listed in Table 1 in the framework of the Spanish Meteor Network (SPMN). Each station employed between 4 and 12 cameras, and the field of view covered by each device ranged from 62x50 to 14x11 degrees, approximately. A detailed description of these systems can be found in (Madiedo & Trigo-Rodríguez 2008; Madiedo et al. 2010; Madiedo et al. 2013a). The reduction of the images recorded by these cameras was performed with the AMALTHEA software (Madiedo et al. 2013b), which was developed by the second author and calculates atmospheric trajectories and meteoroid orbits by following the methods described in Ceplecha (1987).

To obtain meteor emission spectra, holographic diffraction gratings were attached to the lens of some of the above-mentioned Watec cameras. These had 500 or 1000 grooves/mm, depending on the device. The configuration of these slitless videospectrographs is explained in (Madiedo 2014). The NSC spectrum recorded in the framework of this spectroscopic survey was analyzed with the CHIMET software, which was also developed by the second author (Madiedo et al. 2013a).

**3 OBSERVATIONS AND RESULTS**

Optimal weather conditions in the South of Spain during most of the months of May and June favoured the monitoring of meteor activity related to the NSC meteoroid stream between 2010 and 2012. In total, 11 double-station NSC meteor trails were imaged in the time period from May 23 to June 15, with absolute magnitudes M ranging from 1.5 to -8.5. These are listed in Table 2, where their SPMN code is given for identification. The photometric mass $m_p$ of these events ranges from 0.32 to 41 g. As in previous papers (see e.g. Madiedo et al. 2014a), this mass has been calculated from the lightcurve of each event and by using the luminous efficiency given by Ceplecha and McCrosky (1976). Ten additional single-station meteors with good alignment with the position of the NSC radiant were recorded, although these were not taken into consideration since no atmospheric trajectory or orbital data could be derived from them. For meteor SPMN060610, whose peak brightness was equivalent to mag. -8.5 ± 0.5, the emission spectrum was recorded by one videospectrograph located at station #5 in Table 1 (Sierra Nevada). Table 2 also shows the main parameters of the atmospheric path of these double-station meteors: the beginning and ending height of the meteor ($H_b$ and $H_e$, respectively), the pre-atmospheric velocity ($V_\infty$) the geocentric velocity ($V_g$). The right ascension $\alpha_g$ and declination $\delta_g$ of the geocentric radiant (J2000) are also listed. The orbital parameters derived for the progenitor meteoroids are included in Table 3. This table also lists the calculated value of the Tisserand parameter $T_J$ and the orbital period P of these particles. The association of these meteoroids with the NSC stream, whose orbit was taken from Sekanina (1976) and is also listed in Table 3, was performed on the basis of the Southworth & Hawkins $D_{SH}$ criterion (Southworth & Hawkins 1963). This criterion remains below the usually accepted cut-off value of 0.15 (Lindblad 1971a,b). Table 3 also shows the orbital elements of Asteroid 1996 JG, the potential parent body of the stream. The aphelion distance of this asteroid is Q = 2.99 AU (Jenniskens 2006).

One emission spectrum resulted from our spectroscopic survey. Thus, on June 6, 2010 a videospectrograph operating at Sierra Nevada imaged the emission spectrum of the mag. -8.5 ± 0.5 NSC meteor with code 060610. The spectral response of this device is shown in Figure 3 in (Madiedo et al. 2014b). As in previous works, the signal was processed with the CHIMET by following the procedure explained in (Madiedo et al. 2013a). The emission lines produced by the Na doublet at 588.9 nm and the Mg triplet at 516.7 nm were employed to calibrate this spectrum in wavelength by using typical metal lines appearing in meteor spectra. This spectrum is shown in Figure 1, which includes the main contributions by following the multiplet numbering proposed by Moore (1945). The observed features are blends of several lines. The most prominent contributions correspond to Mg I-2 (centred at 516.7 nm) and several Fe I multiplets, such as Fe I-5 (centred at 367.9 nm), Fe I-42 (centred at 420.2 nm), Fe I-41 (centred at 441.5 nm) and Fe I-15 (centred at 526.9 nm).

**4 DISCUSSION**

**4.1 Meteoroid strength**

We have estimated the toughness of meteoroids belonging to the NSC stream by analyzing the flares exhibited the meteors produced by these particles. Only two of the meteors discussed in this work exhibited such flares (events 060610 and 270512 in Table 2). These flares occur as a consequence of the sudden fragmentation of the meteoroids in the atmosphere when the aerodynamic pressure becomes larger than the meteoroid strength. This aerodynamic pressure P is given by the following equation:

$$P = \rho_{atm} \cdot v^2 \qquad (1)$$

where v is the velocity of the meteoroid and $\rho_{atm}$ the atmospheric density at the corresponding height. We have obtained the atmospheric density by employing the US standard atmosphere model (U.S. Standard Atmosphere 1976). The flare is produced by the fast ablation of tiny fragments produced during this fragmentation, which are delivered to the thermal wave in the fireball's bow shock. According to this approach, the critical pressure under which the fragmentation takes place can be used as an estimation of the tensile strength of the meteoroid (Trigo-Rodriguez & Llorca 2006).

Table 4 shows the tensile strength estimated from Eq. (1) for the parent meteoroids of events 060610 and 270512, and also the corresponding meteor velocity and break-up height. According to this analysis, the strength of NSC meteoroids ranges from $(2.6 \pm 0.2) \cdot 10^{-2}$ MPa for the 270512 event to $(1.1 \pm 0.1) \cdot 10^{-1}$ MPa for the 606010 meteor. These values are lower that the average strength inferred for meteoroids belonging to the Geminid stream $((2.2 \pm 0.2) \cdot 10^{-1}$ MPa) (Trigo-Rodriguez & Llorca 2006) whose generally accepted parent body is asteroid (3200) Phaethon. But when compared with cometary streams with a similar entry velocity, these values are similar to the average $(8.5 \pm 0.3) \cdot 10^{-2}$ MPa strength found for the α-Capricornids (Madiedo et al. 2014c), higher than the average $(4.2 \pm 0.3) \cdot 10^{-2}$ MPa strength found for the Camelopardalids (Madiedo et al. 2014d) and also higher than the typical $10^{-3}$ MPa strength found for the October Draconids (Trigo-Rodriguez et al. 2013). So, from the calculated strength values we cannot clearly conclude whether the NSC stream has an asteroidal or a cometary origin.

**4.2 Meteor heights**

Figure 2 can be employed to analyze the dependence of the beginning and terminal heights of the NSC meteors with the mass of their progenitor meteoroid. With respect to the terminal point of the luminous trajectory, the experimental data in this plot reveals that, as expected, $H_e$ decreases with the logarithm of the photometric mass of the meteoroid. However, the initial height $H_b$ is practically constant and remains in the level between 95 and 100 km for the mass range considered here. This situation for $H_b$ is similar to that found for the asteroidal Geminid meteoroids, and different to the behaviour exhibited by cometary meteoroids, where the initial height increases with increasing meteoroid mass (Jenniskens 2004, Madiedo 2015). This suggests an asteroidal origin for the NSC meteoroid stream, as is also deduced from the averaged value found for the Tisserand parameter (Table 3), which yields $T_J = 3.52 \pm 0.08$. It is worth mentioning that comets are the parent bodies of most meteoroid streams (Jenniskens 2006). Thus, in almost all cases in which a generally accepted association between a meteoroid stream and a parent body has been established, the progenitor object is an active comet (Jopek et al. 2002; Jenniskens 2006). The only significant exception is the Geminid stream, whose parent is (3200) Phaeton, which however is considered by some researchers as an extinct cometary nucleus rather than a regular asteroid (Čapek & Borovička 2009). Despite establishing a link between the NSC stream and asteroid 1996 JG is out of the scope of this work, the data analyzed here confirm the likely asteroidal nature of these meteoroids.

### 4.3 Emission spectrum

A remarkable feature of the 060610 spectrum is the relatively low intensity of the Na I-1 line in relation to the Mg I-2 emission. The likely reason for this is the depletion of sodium in the progenitor meteoroid with respect to the expected value for chondritic materials. To confirm this, the relative intensities of the emission lines of multiplets Na I-1, Mg I-2 and Fe I-15 were obtained, since these can provide information about the nature of this meteoroid (Borovička et al. 2005). These intensities were obtained with CHIMET. This software measured and added the brightness of these lines frame by frame in the videospectrum. Then, the time-integrated intensities for each multiplet were corrected by taking into account the spectral sensitivity curve of the spectrograph. This calculation provided the following intensity ratios: Na/Mg = 0.47 and Fe/Mg = 0.80. As expected, the Na/Mg intensity quotient does not fit the expected value for meteoroids with chondritic composition for a meteor velocity of ~ 22 km s$^{-1}$, as Figure 5 in Borovička et al. (2005) shows. The depletion in Na is clearly seen in the ternary diagram shown in Figure 3. This plot shows the expected relative intensity of the Na I-1, Mg I-2 and Fe I-15 lines for chondritic materials as a function of the velocity of the meteor (solid line) (Borovička et al. 2005). The cross in this diagram shows the value measured for the 060610 spectrum, which significantly deviates from the chondritic value for a meteor velocity of around 22 km s$^{-1}$. Since the perihelion distance is too large (q = 0.639 ± 0.004 AU) to explain this depletion on the basis of a close approach to the Sun, this composition is most likely related to a non-chondritic nature of the parent body. However, it cannot be discarded that as a consequence of orbital evolution the parent object of this meteoroid had a significantly lower q value in the past.

### 5 CONCLUSIONS

The monitoring of the night sky during the activity period of the North ω-Scorpiids provided 11 double-station NSC events and the emission spectrum of one member of this shower. The Tisserand parameter with respect to Jupiter and the constancy of the initial height of the luminous trajectory as the mass of the meteoroid increases suggests that this stream has an asteroidal origin.

The emission spectrum recorded for a mag. -8.5 NSC meteor exhibited as main contributions the emission lines of Mg I-2 and several Fe I multiplets. The Na I-1 line, however, was dimmer than expected for meteoroids with a chondritic composition. The analysis of the relative intensities of Mg I-2, Fe I-15 and Na I-1 in this signal confirmed a depletion of Na with respect to the chondritic value. The perihelion distance cannot explain this depletion on the basis of a close approach to the Sun and so this feature may be related to a non-chondritic composition of the progenitor asteroid.

## ACKNOWLEDGEMENTS

The meteor observing stations involved in this research were funded by J.M. Madiedo. We thank AstroHita Foundation for its continuous support in the operation of the meteor observing station located at La Hita Astronomical Observatory. We acknowledge Dr. Alberto J. Castro-Tirado and Dr. José L. Ortiz for their support in the operation of the meteor observing station at Sierra Nevada Observatory. We also thank Dr. Jiri Borovička and an anonymous referee for their revision of this paper.

**TABLES**

Table 1. Geographical coordinates of the meteor observing stations involved in this work.

| Station # | Station name | Longitude (W) | Latitude (N) | Altitude (m) |
|---|---|---|---|---|
| 1 | Sevilla | 5º 58' 50" | 37º 20' 46" | 28 |
| 2 | Cerro Negro | 6º 19' 35" | 37º 40' 19" | 470 |
| 3 | El Arenosillo | 6º 43' 58" | 37º 06' 16" | 40 |
| 4 | Huelva | 6º 56' 11" | 37º 15' 10" | 25 |
| 5 | Observatorio de Sierra Nevada (OSN) | 3º 23' 05" | 37º 03' 51" | 2896 |
| 6 | La Hita | 3º 11' 00" | 39º 34' 06" | 674 |

Table 2. Atmospheric trajectory and radiant data (J2000) for the 11 NSC meteor events discussed here (M: absolute magnitude; $m_p$: photometric mass; $H_b$ and $H_e$: beginning and ending height of the luminous phase, respectively; $\alpha_g$, $\delta_g$: right ascension and declination of the geocentric radiant; Z: zenith distance of the apparent radiant; $V_\infty$, $V_g$: observed preatmospheric, and geocentric velocities, respectively).

| Meteor code | Date and Time (UTC) ±0.1s | M ±0.5 | $m_p$ (g) | $H_b$ (km) | $H_e$ (km) | $\alpha_g$ (º) | $\delta_g$ (º) | Z (º) | $V_\infty$ (km s$^{-1}$) | $V_g$ (km s$^{-1}$) |
|---|---|---|---|---|---|---|---|---|---|---|
| 010610b | 1 June 2010 2h54m13.7s | -0.5 | 2.2±0.3 | 97.3 | 77.5 | 246.71±0.12 | -12.0±0.1 | 57.8 | 22.1±0.3 | 19.3±0.3 |
| 020610a | 2 June 2010 1h00m38.1s | 1.0 | 0.67±0.07 | 96.2 | 82.7 | 248.47±0.10 | -17.0±0.1 | 50.5 | 23.0±0.3 | 20.2±0.3 |
| 020610b | 2 June 2010 1h38m47.6s | 1.5 | 0.41±0.04 | 94.5 | 83.2 | 246.78±0.08 | -11.9±0.1 | 49.7 | 22.2±0.3 | 19.4±0.3 |
| 030610 | 3 June 2010 0h23m19.2s | -2.0 | 5.1±0.5 | 95.9 | 75.3 | 249.01±0.09 | -11.8±0.2 | 45.8 | 22.7±0.3 | 19.8±0.3 |
| 060610 | 6 June 2010 23h18m42.0s | -8.5 | 41±5 | 98.5 | 50.7 | 251.76±0.09 | -10.9±0.1 | 42.5 | 22.1±0.4 | 19.0±0.4 |
| 070610a | 7 June 2010 3h05m21.8s | -5.0 | 9.9±1.2 | 97.6 | 71.2 | 255.81±0.13 | -16.8±0.1 | 61.3 | 23.8±0.4 | 21.2±0.4 |
| 070610b | 7 June 2010 5h17m21.5s | -3.0 | 5.8±0.6 | 99.5 | 76.3 | 253.26±0.23 | -17.7±0.2 | 82.0 | 22.6±0.4 | 20.0±0.4 |
| 010611 | 1 June 2011 0h50m57.4s | 1.5 | 0.32±0.03 | 99.3 | 87.6 | 248.45±0.10 | -11.3±0.1 | 45.6 | 22.7±0.3 | 19.8±0.3 |
| 050611 | 5 June 2011 23h41m02.8s | 0.5 | 0.81±0.08 | 98.4 | 80.2 | 250.35±0.11 | -15.8±0.2 | 50.1 | 22.4±0.4 | 19.4±0.5 |
| 270512 | 27 May 2012 0h03m42.1s | -6.0 | 15.2±1.8 | 96.3 | 72.6 | 246.31±0.06 | -13.0±0.1 | 47.2 | 24.1±0.4 | 21.3±0.4 |
| 300512 | 30 May 2012 1h19m00.7s | -2.5 | 6.1±0.7 | 95.2 | 78.0 | 243.74±0.08 | -12.5±0.2 | 49.3 | 22.1±0.4 | 19.2±0.4 |

Table 3. Orbital elements (J2000) for the meteors discussed in the text. The orbit of Asteroid 1996 JG (Jenniskens 2006) and the nominal orbit for the NSC stream (Sekanina 1976) have been also indicated. The Tisserand parameter with respect to Jupiter ($T_J$) and the value of the $D_{SH}$ Southworth and Hawkins criterion are also listed.

| Object | a (AU) | e | i (°) | Ω (°) ± $10^{-5}$ | ω (°) | q (AU) | P (yr) | $T_J$ | $D_{SH}$ |
|---|---|---|---|---|---|---|---|---|---|
| 010610  | 1.96±0.05 | 0.665±0.011 | 5.9±0.2 | 70.37943 | 263.17±0.18 | 0.657±0.004 | 2.75 | 3.56±0.06 | 0.02 |
| 020610a | 1.98±0.06 | 0.685±0.011 | 3.2±0.1 | 71.25175 | 266.95±0.13 | 0.623±0.003 | 2.79 | 3.52±0.07 | 0.02 |
| 020610b | 2.04±0.06 | 0.673±0.012 | 6.0±0.2 | 71.28754 | 261.50±0.13 | 0.666±0.004 | 2.91 | 3.47±0.07 | 0.03 |
| 030610  | 2.01±0.06 | 0.677±0.012 | 6.6±0.2 | 72.19630 | 263.79±0.18 | 0.649±0.003 | 2.85 | 3.49±0.07 | 0.02 |
| 060610  | 1.97±0.08 | 0.659±0.017 | 6.9±0.2 | 75.98340 | 261.36±0.19 | 0.672±0.004 | 2.78 | 3.55±0.09 | 0.07 |
| 070610a | 1.99±0.07 | 0.704±0.014 | 4.3±0.2 | 76.12725 | 271.02±0.20 | 0.588±0.005 | 2.81 | 3.49±0.08 | 0.02 |
| 070610b | 1.98±0.06 | 0.684±0.014 | 3.2±0.2 | 76.20891 | 266.39±0.36 | 0.628±0.007 | 2.81 | 3.51±0.07 | 0.02 |
| 010611  | 1.88±0.05 | 0.664±0.012 | 6.9±0.2 | 70.05239 | 266.62±0.18 | 0.633±0.004 | 2.59 | 3.65±0.07 | 0.03 |
| 050611  | 2.00±0.09 | 0.673±0.017 | 4.0±0.2 | 74.78615 | 263.13±0.23 | 0.655±0.005 | 2.84 | 3.51±0.09 | 0.03 |
| 270512  | 1.93±0.08 | 0.697±0.015 | 6.2±0.2 | 65.93607 | 271.47±0.18 | 0.586±0.004 | 2.69 | 3.55±0.09 | 0.03 |
| 300512  | 2.02±0.09 | 0.672±0.017 | 5.3±0.2 | 68.86180 | 260.85±0.12 | 0.671±0.005 | 2.93 | 3.46±0.09 | 0.04 |
| NSC     | 1.943 | 0.671 | 6.8 | 65.9 | 265.4 | 0.639 | 2.70 | 3.57 | - |
| 1996JG  | 1.802 | 0.661 | 5.2 | 63.370 | 269.733 | 0.610 | 2.42 | 3.77 | - |

Table 4. Aerodynamic pressure for the flare exhibited by NSC meteors.

| Meteor code | Height (km) | Velocity (km s$^{-1}$) | Aerodynamic pressure (MPa) |
|---|---|---|---|
| 060610 | 61.5±0.5 | 21.5±0.5 | (1.1±0.1)·$10^{-1}$ |
| 270512 | 73.4±0.5 | 23.9±0.5 | (2.6±0.2)·$10^{-2}$ |

**FIGURES**

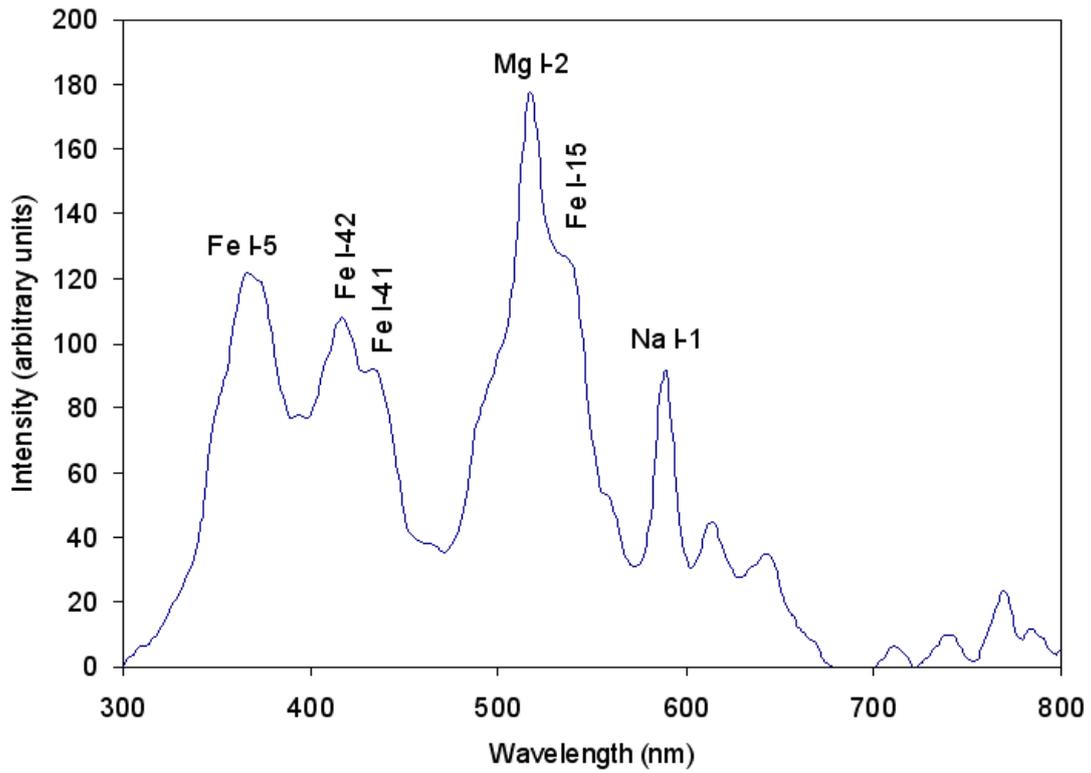

Figure 1. Calibrated emission spectrum of the 060610 NSC meteor, integrated along the atmospheric path of the event. The most relevant lines have been indicated.

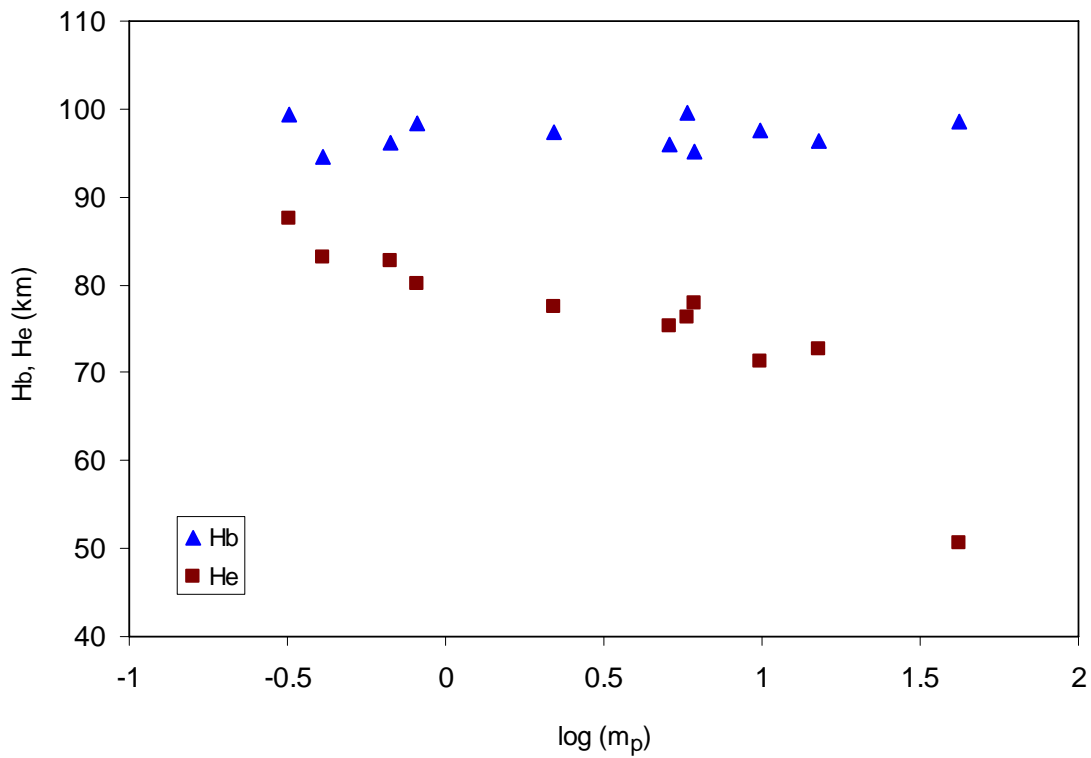

Figure 2. Meteor beginning ($H_b$) and terminal ($H_e$) heights vs. logarithm of the photometric mass $m_p$ of the meteoroid.

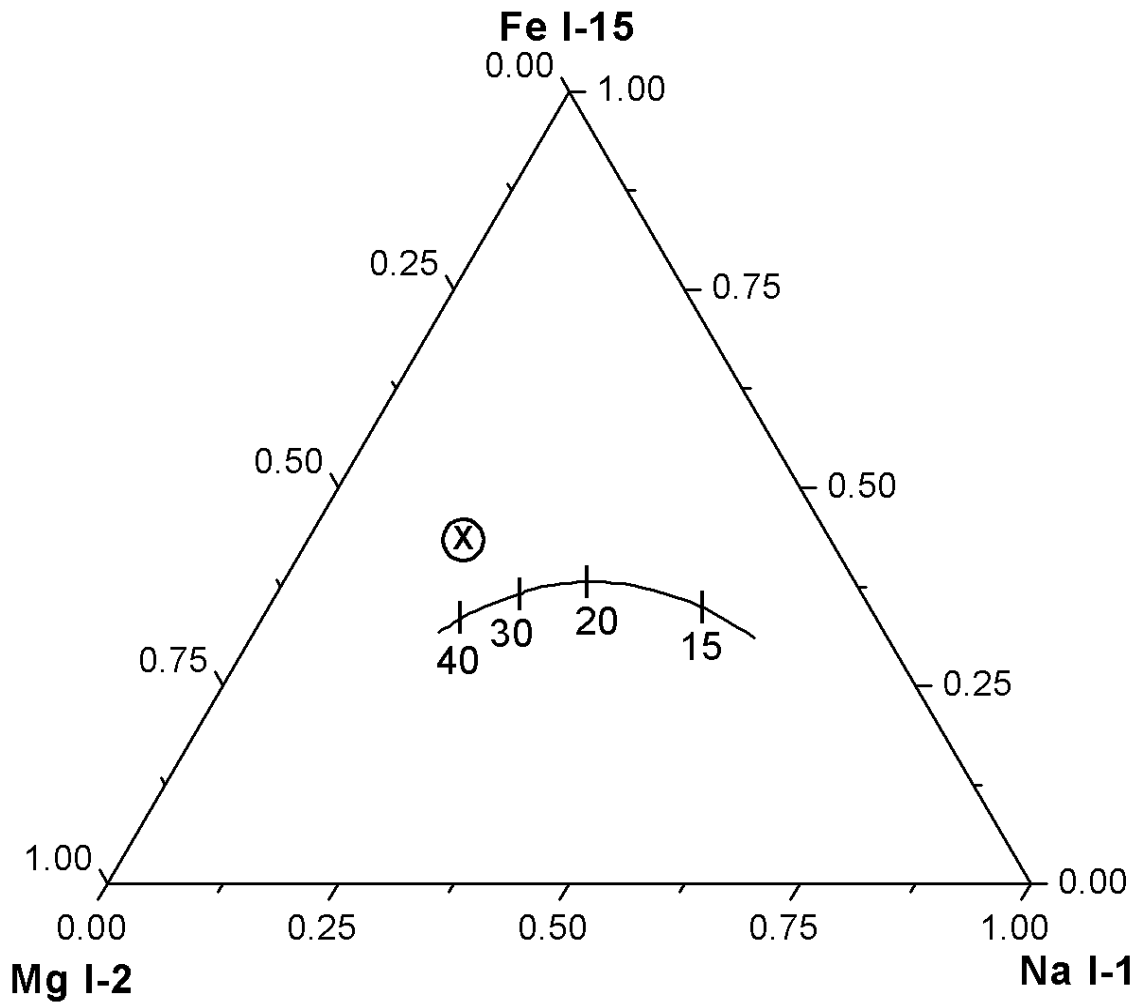

Figure 3. Expected relative intensity (solid line), as a function of meteor velocity (in km s$^{-1}$), of the Na I-1, Mg I-2 and Fe I-15 multiplets for chondritic meteoroids (Borovička et al., 2005). Cross: experimental relative intensity obtained for the 060610 NSC meteor; circle around cross: uncertainty (error bars) for this experimental value.